\documentclass[10pt]{elsart}%
\usepackage{epsfig}
\usepackage{graphicx}
\usepackage{amssymb}
\usepackage{amsmath}
\usepackage{url}
\usepackage{accents}
\usepackage{amsfonts}%
\providecommand{\U}[1]{\protect\rule{.1in}{.1in}}
\begin{document}
\begin{frontmatter}
\title{Unified thermopower in the variable range hopping regime}
\author{Said Boutiche\corauthref{email}}
%\author{Said Boutiche}
\thanks[email]{E-mail: vizuallearning@hotmail.com}
\address{Dept. de Physique, Universite de Bechar. 08000- Bechar. Algeria}
\begin{abstract}
Since nearly 4 decades, various theoretical behaviours have been found for the thermopower
in the variable range hopping regime. In 1969, Cutler and Mott have predicted a linear
variation with temperature T of the thermopower: S = const.T. In the seventies, it has been
found by Zvyagin, Overhof and Mott that S = const.T(1/2). In 1986, Triberis and Friedman have
found S = const.T(-1/4) . But there is up to now no theoretical formulation of the thermopower
when this one is T-independent. By choosing a specific distribution for the density of states, we
show in this paper that all behaviours above can be unified in a unique thermopower formula. We
find in addition with this formula, a T-independent expression given by: S=(L/xi)(k/e), in which
xi is the wave function decay length and L is a characteristic length, depending on the form of the
density of states.
\end{abstract}
\end{frontmatter}

%main text

\section{Introduction}

Mott's paper1 \cite{1}\ of 1969 has played a key role in the construction of
the important theoretical edifice \cite{2}-\cite{9} on which is rested our
understanding of the hopping conduction mechanism. The electric properties of
various disordered systems and amorphous semiconductors have been explained
\cite{10} during more than 3 decades by using Mott's variable range hopping
(VRH) theory. Such a theory is still extensively used today in varied fields,
to investigate the mechanism of charge transport for example as well in
correlated electron systems \cite{11} as in biological systems as DNA
\cite{12}.

However, the study in the VRH regime of some thermoelectric properties such as
the thermopower have been somewhat neglected. It seems indeed, that this one
is ill adapted in its present form to support the experimental measurements of
the Seebeck coefficient carried out on different VRH systems. A simple
illustration of this is its persistent disobedience during more than 3
decades, to the predicted theoretical laws \cite{4}-\cite{8}, observed in most
amorphous semiconductors \cite{6}, \cite{8}, \cite{13}-\cite{17}.

Because of new environmental problems, the field of the thermoelectric power
generation has attracted recently a major interest in materials with high
thermoelectric properties \cite{18}. In this context, the explanation of the
experimental behaviour of the VRH thermopower remains still problematic since
there is not yet a clear theory that explains some TEP behaviours when
conduction is by variable range hopping, as the divergent one observed at low
temperature in ref \cite{19} or the temperature independent one, reported by
the authors of ref \cite{20}.

Such a theoretical insufficiency of the TEP is attributed in this work to the
fragility of the original hypothesis concerning the nature of the density of
states (DOS), with which the evaluation of the Seebeck coefficient has been
made. By reconsidering the hypothesis of the "slow linear variation of the
DOS" near Fermi level on which most of the VRH theories of the TEP are rested
\cite{4}-\cite{8}, we investigate by using the percolation theory both
conductivity and thermoelectric power when the DOS takes the asymmetric
generalized form:%

\begin{equation}
N(E)=N(E_{F})+g(E) \label{eq.1}%
\end{equation}
where N(E$_{F}$) is a finite density of states at Fermi level E$_{F}$ and g(E)
represents the asymmetrical part of N(E):%

\begin{equation}
g(-E)=-g(E) \label{eq.2}%
\end{equation}

This choice of g(E), is motivated by the following reason: by using the
concept of the random network of conductances \cite{2},\cite{3},\cite{6}, we
can expect to find Mott T$^{-1/4}$ conductivity again since the number of
conductances that belong to the critical path \cite{3},\cite{6}, generated by
N(E) of eq.(\ref{eq.1} -\ref{eq.2})would be quasi similar to the one generated
by a constant DOS \cite{2}. If so, we can expect to obtain different
thermopower classes that correspond to the Mott T$^{-1/4}$ conductivity,
rather than the unique an insufficient formula obtained by the classic VRH theories.

\section{Percolation method for the VRH problem}

It has been shown by using the percolation theory \cite{2}-\cite{7}, that the
conduction problem between localised states is equivalent to the conduction
problem through a random network of conductances. Each conductance
$\sigma_{ij}$ defined by:%

\begin{equation}
\sigma_{ij}=\exp\{-\frac{2r_{ij}}{\xi}-\frac{\left\vert E_{i}\right\vert
+\left\vert E_{j}\right\vert +\left\vert E_{i}-E_{j}\right\vert }{2kT}\}
\label{eq.3}%
\end{equation}
links two sites located at energies E$_{i}$ and E$_{j}$, separated in space by
the distance r$_{ij}$. In eq.(\ref{eq.3}), k is the Boltzmann constant, T is
the temperature and $\xi$ is the decay length of the wave function.

To solve a conduction-percolation problem, two critical conditions must be
satisfied: the first concerns the number m(E$_{i}$) of incoming conductances
(or bonds) to a site located at energy E$_{i}$ and positioned at the center of
a sphere of radius r$_{ij}$ :%

\begin{equation}
m(E_{i})=\frac{4\pi}{3}\int r_{ij}^{3}N(E_{j})dE_{j} \label{eq.4}%
\end{equation}
When averaged by a weighting probability function, m(E$_{i}$) must reach the
critical concentration c of conductances per site, solicited for conduction
within an active energy layer $\Delta$, given by (percolation criterion):%

\begin{equation}
c=\frac{%
%TCIMACRO{\dint \limits_{-\Delta}^{\Delta}}%
%BeginExpansion
{\displaystyle\int\limits_{-\Delta}^{\Delta}}
%EndExpansion
m^{2}(E_{i})N(E_{i})dE_{i}}{%
%TCIMACRO{\dint \limits_{-\Delta}^{\Delta}}%
%BeginExpansion
{\displaystyle\int\limits_{-\Delta}^{\Delta}}
%EndExpansion
m(E_{i})N(E_{i})dE_{i}} \label{eq.5}%
\end{equation}
The second critical condition concerns the nature of conductances that are
solicited for conduction. To be an efficient conductor, each conductance
$\sigma_{ij}$ must be larger than a critical conductance given by:%

\begin{equation}
\sigma_{c}=\exp\{-\frac{\Delta}{kT}\} \label{eq.6}%
\end{equation}
Physically, when these two critical conditions are satisfied, it appears a
continuous (critical) path of conductances, joining one side of the VRH system
to the other. The problem of the random network is said solved only when
$\sigma_{c}$ is identified, and this occurs when eq. (\ref{eq.5}) is solved
with respect to $\Delta$.

\section{Conductivity of the asymmetric DOS}

We show now that our prediction to find again Mott T$^{-1/4}$ conductivity for
the DOS given by eq.(\ref{eq.1}, \ref{eq.2}) is true. We take for this a
generalized asymmetric DOS form given by:%

\begin{equation}
N(E)=N(E_{F})+s_{q}.E^{q}=N(E_{F})\left[  1+\nu_{q}.E^{q}\right]
\ \label{eq.7}%
\end{equation}
In this equation s$_{q}$ is a positive constant, the energy E is measured from
the Fermi level E$_{F}$=0 and q is a real number so that g(E) obeys
eq.(\ref{eq.2}). Typical values of q can be q=1, 3, 5, etc, or q=1/3, 1/5, 1/7
or 5/3, 5/7 etc.

Let start our conductivity computation by evaluating the number m(E$_{i}$),
when $\sigma_{ij}$%
%TCIMACRO{\TEXTsymbol{>}}%
%BeginExpansion
$>$%
%EndExpansion
$\sigma_{c}.$ For E$_{i}$
%TCIMACRO{\TEXTsymbol{>}}%
%BeginExpansion
$>$%
%EndExpansion
0 eq.(\ref{eq.4}) yields:\bigskip%

\begin{equation}
m(E_{i})=\frac{4\pi}{3}\left(  \frac{\xi}{2kT}\right)  ^{3}\left[  I_{1}%
+I_{2}+I_{3}\right]  \label{eq.8}%
\end{equation}
where the I$_{1}$, I$_{2}$ and I$_{3}$ integrals are given by:

$I_{1}=\int_{0}^{E_{i}}(\Delta-E_{i})^{3}N(E_{j})dE_{j}$

$I_{2}=\int_{E_{i}}^{\Delta}(\Delta-E_{j})^{3}N(E_{j})dE_{j}$

$I_{3}=\int_{-\Delta+E_{i}}^{0}(\Delta-E_{i}+E_{j})^{3}N(E_{j})dE_{j}$

\begin{figure}[ptbh]
\begin{center}
\includegraphics{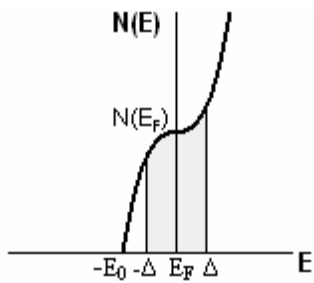}
\end{center}
\caption{ Density of states representation with an active energy layer
$\Delta<E_{0}$. }%
\label{fig.1}%
\end{figure}

By inserting the DOS of eq.(\ref{eq.7}) in I$_{1}$, I$_{2}$ and I$_{3}$ and by
putting x=E$_{i}$/$\Delta,$ a new dimensionless energy variable,
eq.(\ref{eq.8}) yields after an extremely laborious calculation:%

\begin{equation}
m(x,q)=m_{0}(x)+\mu(x,q) \label{eq.9}%
\end{equation}
We have written in eq.(\ref{eq.9}) the number of conductances attached to the
site of energy E$_{i}$ as the sum of two terms: $m_{0}(x)$ represents the
number of conductances resulting from the symmetrical part N(E$_{F}$) of the
density of states and $\mu(x,q)$ is the number of conductances resulting from
the asymmetrical part of N(E):%

\begin{equation}
m_{0}(x)=\frac{M_{0}}{2}(1+x)(1-x)^{3} \label{eq.10}%
\end{equation}

\begin{equation}
\mu(x,q)=M_{0}\frac{\Delta^{q}}{E_{0}^{q}}\frac{6\Gamma(q+1)}{\Gamma
(q+5)}f(x,q) \label{eq.11}%
\end{equation}
Here -E$_{0}$ is an energy below E$_{F}$ (see Fig.\ref{fig.1}), solution of
the equation N(E) = 0; $\Gamma(z)$ is the gamma function of z and the
parameters M$_{0}$ and $f(x,q)$ are given by the following expressions:

$M_{0}=\frac{4\pi}{3}\left(  \frac{\xi}{2kT}\right)  ^{3}\Delta^{4}N(E_{F})$

$f(x,q)=1-A_{q}.x^{q+2}+B_{q}.x^{q+3}-C_{q}.x^{q+4}-(1-x)^{q+4}$

where we have: $A_{q}=(q^{2}+7q+12)/2;$ $B_{q}=q^{2}+6q+8;$ $C_{q}%
=(q^{2}+5q+6)/2$

We are now ready to solve our percolation-conductivity problem by replacing
eq.(\ref{eq.7}) and eq.(\ref{eq.9}) in eq.(\ref{eq.5}):%

\begin{equation}
c=\frac{%
%TCIMACRO{\dint \limits_{-1}^{1}}%
%BeginExpansion
{\displaystyle\int\limits_{-1}^{1}}
%EndExpansion
\left[  m_{0}^{2}(x)+2m_{0}(x)\mu(x,q).\nu_{q}.x^{q}+\mu^{2}(x,q)\right]  dx}{%
%TCIMACRO{\dint \limits_{-1}^{1}}%
%BeginExpansion
{\displaystyle\int\limits_{-1}^{1}}
%EndExpansion
\left[  m_{0}(x)+\mu(x,q).\nu_{q}.x^{q}\right]  dx} \label{eq.12}%
\end{equation}

In eq.(\ref{eq.12}) we have omitted to write the asymmetrical functions since
their integrations cancel over positive and negative energies.

\begin{figure}[h]
\begin{center}
\includegraphics{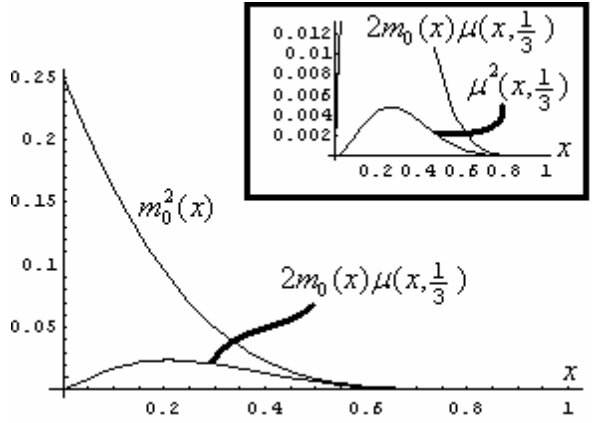}
\end{center}
\caption{Representation of two integrand terms of eq.(\ref{eq.12}) in unit of
$M_{0}^{2}$ for q=1/3. The inset figure shows how much $\mu^{2}(x,q)$ in
eq.(\ref{eq.12}) is small.}%
\label{fig.2}%
\end{figure}

Before tackling the tedious resolution of this equation, may be it would be
advantageous to examine and compare first the terms of each integrand. If we
place ourselves in the situation shown in fig.\ref{fig.1}, where the
temperature of the VRH system is such as the active energy layer $\Delta
<E_{0}$ , it becomes then apparent that we can neglect in eq.(\ref{eq.12}) all
terms containing the $\mu(x,q)$ number since the surface delimited by these
terms is negligible in comparison with the one corresponding to $m_{0}^{2}(x)$
when q$\geq1/3$\ (see Fig.\ref{fig.2}).

\bigskip

In such a situation, eq.(\ref{eq.12}) becomes quasi similar to the one found
in \cite{3} for a constant DOS:%

\begin{equation}
c=\frac{%
%TCIMACRO{\dint \limits_{-1}^{1}}%
%BeginExpansion
{\displaystyle\int\limits_{-1}^{1}}
%EndExpansion
m_{0}^{2}(x)dx}{%
%TCIMACRO{\dint \limits_{-1}^{1}}%
%BeginExpansion
{\displaystyle\int\limits_{-1}^{1}}
%EndExpansion
m_{0}(x)dx} \label{eq.13}%
\end{equation}
Integrating eq.(\ref{eq.13}) and taking as in \cite{3} c = 1.7, we
obtain:\bigskip%

\begin{equation}
\frac{\Delta}{kT}=\left(  \frac{T_{0}}{T}\right)  ^{\frac{1}{4}} \label{eq.14}%
\end{equation}
where T$_{0}^{1/4}=1.8/(k.N(E_{F}).\xi^{3})^{1/4}.$The numerical value 1.8 of
T$_{0}^{1/4}$ is similar to the one found earlier by Pollak in \cite{3}.
Consequently, equations (\ref{eq.6}) and (\ref{eq.14}) show that Mott
T$^{-1/4}$ conductivity is valid with a very good approximation for any DOS of
eq.(\ref{eq.7}), when q$\geq1/3$.

In the limit q$\rightarrow$0 (step like DOS at E$_{F}$), we lose the
approximation that gave eq.(\ref{eq.13}) since $m_{0}^{2}(x)$ and
2$m_{0}(x)\mu(x,0)$ become comparable. The re-evaluation of eq.(\ref{eq.12})
in this case yields:

\begin{center}%
\begin{equation}
c\approx\frac{%
%TCIMACRO{\dint \limits_{-1}^{1}}%
%BeginExpansion
{\displaystyle\int\limits_{-1}^{1}}
%EndExpansion
\left[  m_{0}^{2}(x)+2m_{0}(x)\mu(x,0).\nu_{0}\right]  dx}{%
%TCIMACRO{\dint \limits_{-1}^{1}}%
%BeginExpansion
{\displaystyle\int\limits_{-1}^{1}}
%EndExpansion
\left[  m_{0}(x)+\mu(x,0).\nu_{0}\right]  dx} \label{eq.15}%
\end{equation}

$=\frac{4\pi}{3}\frac{85}{252}\left(  \frac{\xi}{2kT}\right)  ^{3}%
.N(E_{F}).\Delta^{4}$
\end{center}

Here again the conductivity follows Mott law: $\sigma=\exp\{-(T_{0}%
/T)^{-1/4}\}$ in which the new numerical constant of T$_{0}^{-1/4}$ takes the
value: 1.76. Such a negligible change of T$_{0}^{-1/4}$ is of course
indetectable during any experimental mesurement, implying the validity of
eq.(\ref{eq.14}) for any q.

\section{The unified thermopower}

To compute the Seebeck coefficient S, we make use of the formula:%

\begin{equation}
S=\frac{k}{e}\frac{\Pi}{kT} \label{eq.16}%
\end{equation}
here e is the electronic charge and $\Pi$ is the Peltier heat. Equation
\ref{eq.16} is one of the classical Kelvin relations of thermoelectricity. For
metals it gives \cite{21}:%

\begin{equation}
S=\frac{\pi^{2}}{3}\frac{k}{e}kT\left[  \frac{\partial\ln\sigma(E)}{\partial
E}\right]  _{E=E_{F}} \label{eq.17}%
\end{equation}

In the variable range hopping regime, $\Pi$ may be interpreted \cite{5} as the
mean energy of any site that belongs to the continuous critical path of
conductances. We can then write it under the form:%

\begin{equation}
\Pi=\frac{\int E.N(E).p(E)dE}{\int N(E).p(E)dE} \label{eq.18}%
\end{equation}

The Peltier heat in this expression appears as an energy E, weighted the by a
probability factor p(E) that is assumed proportional to the number m(E) of
conductances attached to a site located at energy E. Inserting equations
(\ref{eq.7}) and (\ref{eq.9}) in eq.(\ref{eq.18}) we obtain:%

\begin{equation}
\Pi=\frac{\int\left[  N(E_{F}).E.\mu(E,q)+s_{q}.m_{0}(E).E^{q+1}\right]
dE}{\int\left[  N(E_{F}).m_{0}(E)+s_{q}.E^{q}.\mu(E,q)\right]  dE}
\label{eq.19}%
\end{equation}

In this expression, we have omitted again to write the asymmetrical factors of
E since their integrations cancel over positive and negative energies. For
q$\geq1/3$\ we can also neglect the second denominator term since the active
layer of states solicited for the conduction process must verify eqs (1) and
(2), in which case we have $s_{q}.E^{q}<$ $N(E_{F})$. Integrating and
inserting the Peltier heat of eq.(\ref{eq.19}) in the thermopower expression
of eq.(\ref{eq.16}), we obtain:%

\begin{equation}
S=F(q)\frac{k}{e}\frac{\Delta^{q+1}}{kT}\left[  \frac{d\ln N(E)}{dE^{q}%
}\right]  _{E=E_{F}} \label{eq.20}%
\end{equation}
where $F(q)=60\left[  \Gamma(q+1)/\Gamma(q+7)\right]  (q+1)(q+4).$ Replacing
the active layer energy $\Delta$ by its expression of eq.(\ref{eq.14}), we
find that the T dependence of the variable range hopping thermopower is a
class given by:%

\begin{equation}
S=F(q)\frac{k}{e}k^{q}T_{0}^{\frac{q+1}{4}}T^{\frac{3q-1}{4}}\left[
\frac{d\ln N(E)}{dE^{q}}\right]  _{E=E_{F}} \label{eq.21}%
\end{equation}

In the case of a linear asymmetry of density of states (q = 1), the classical
variable range hopping thermopower formula [4-6] is recovered:%

\begin{equation}
S=F(1)\frac{k}{e}k\left(  T_{0}T\right)  ^{\frac{1}{2}}\left[  \frac{d\ln
N(E)}{dE}\right]  _{E=E_{F}} \label{eq.22}%
\end{equation}
with the exact numerical factor found in \cite{5}: F(1) =$\frac{5}{42}$.

In the limit q$\rightarrow\infty$ (constant DOS near Ef), we have S=0 since
F($\infty$)=0. This is in agreement with all thermopower theories indicating
that S=0 when the density of states is symmetric with respect to Fermi level.

Eq.(\ref{eq.21}) is in fact an unified formulation of the thermopower
corresponding to Mott conductivity since it includes according to q, the
following thermopower class/behaviors:

\subsection{Integer metallic class}

It is a class obtained from eq.(\ref{eq.21}) for any odd integer q. S$\sim
T^{1/2}$ (q= 1) is only one element of this class.

\subsection{Fractional metallic class}

It is a class obtained from eq.(\ref{eq.21}) for any fractional q%
%TCIMACRO{\TEXTsymbol{>}}%
%BeginExpansion
$>$%
%EndExpansion
1/3, satisfying eqs.(\ref{eq.1} and \ref{eq.2}). We find in this class for
q=5/3, the first VRH thermopower, predicted in 1969 by Cutler and Mott
\cite{22}.

\subsection{Semiconductor class}

It is a class obtained from eq.(\ref{eq.21}) for any fractional ) with 0%
%TCIMACRO{\TEXTsymbol{<}}%
%BeginExpansion
$<$%
%EndExpansion
q%
%TCIMACRO{\TEXTsymbol{<}}%
%BeginExpansion
$<$%
%EndExpansion
1/3. If q tends to zero as 1/p, where p is an infinite odd number (step like
DOS at E$_{F}$), we find , the thermopower due to small polarons found earlier
by Triberis and Friedman \cite{23} for constant DOS band above E$_{F}$.

\subsection{The constant thermopower}

The pivotal value q=1/3, delimits the metallic and semiconductor classes. Its
thermopower is T-independent and is given by:%

\begin{equation}
S=F(\frac{1}{3})\frac{k}{e}\left(  kT_{0}\right)  ^{\frac{1}{3}}\left[
\frac{d\ln N(E)}{dE^{1/3}}\right]  _{E=E_{F}} \label{eq.23}%
\end{equation}
Putting in this expression $\left[  d\ln N(E)/dE^{1/3}\right]  _{E_{F}%
}=1/E_{0}^{1/3}$ and replacing T$_{0}$ by its expression of eq.(\ref{eq.14}),
we obtain a T-independent formula of the thermopower:%

\begin{equation}
S=\beta\frac{L}{\xi}\frac{k}{e} \label{eq.24}%
\end{equation}
where $\beta$ is a numerical factor of order 1\bigskip\ and :%

\begin{equation}
L=\left(  \frac{1}{N(E_{F}).E_{0}}\right)  ^{\frac{1}{3}} \label{eq.25}%
\end{equation}
is a characteristic length defined by the nature of the DOS. In amorphous
semiconductors where a constant thermopower of magnitude $\approx100\mu
VK^{-1}$can be observed \bigskip\cite{14}, we expect a length of order $\xi$
for L.

\section{Conclusion}

We have shown in this work that the thermopower in the VRH regime may have
according to the DOS shape of eq.(\ref{eq.7}) 3 different behaviours: it can
have a metallic behaviour if q%
%TCIMACRO{\TEXTsymbol{>}}%
%BeginExpansion
$>$%
%EndExpansion
1/3, it can have a semiconductor behaviour if 0%
%TCIMACRO{\TEXTsymbol{<}}%
%BeginExpansion
$<$%
%EndExpansion
q%
%TCIMACRO{\TEXTsymbol{<}}%
%BeginExpansion
$<$%
%EndExpansion
1/3, or it can be T-independent if q=1/3.

\label{}

%The Appendices part is started with the command \appendix;
%appendix sections are then done as normal sections
%\appendix

%\section{}
%\label{}


\begin{thebibliography}{99}                                                                                               %


\bibitem {1}N.F. Mott, Phil. Mag. \textbf{19}, 835 (1969)

\bibitem {2}V. Ambegaokar, B.I. Halperin, and S. Langer, Phys.Rev. \textbf{B
4}, 2612 (1971)

\bibitem {3}M. Pollak, Journ. Non-cryst. Sol. \textbf{11}, 1 (1972)

\bibitem {4}I.P. Zvyagin, phys. stat. sol. (b) \textbf{58}, 443 (1973)

\bibitem {5}I.P. Zvyagin in \textit{Hopping transport in solids,} eds. M.
Pollak and B. I. Shklovskii, (North Holland, Amsterdam, 1991), p.143

\bibitem {6}H. Overhof, phys. stat. sol. (b) \textbf{67}, 709 (1975)

\bibitem {7}M. Pollak, and L. Friedman, \textit{Localization and
Metal-Insulator Transitions} Vol.2 Eds. H. Fritzsche and D. Adler, (Plenum
Press, New-York, 1985), p.347

\bibitem {8}P. Nagels, M. Rotti, and R. Gevers, Journ. Non-cryst. Sol.
\textbf{59-60}, 65 (1983)

\bibitem {9}D. Janaa and J. Fort: Physica B \textbf{344}, 62 (2004)

\bibitem {10}N.F. Mott and E.A. Davis, \textit{Electronic Processes in
Non-Crystalline Materials, 2nd Ed}. (Clarendon Press, Oxford 1979)

\bibitem {11}A. A. Taskin and Yoichi Ando: Phys. Rev. Lett.\textbf{\ 95},
176603 (2005)

\bibitem {12}Z. G. Yu and X. Song, Phys. Rev. Lett.\textbf{\ 86,} 6018 (2001)

\bibitem {13}W. Beyer, and J. Stuke, \textit{Proceedings of the 5th Int. Conf.
on Am. and Liquid Semiconductors}, Eds. Stuke, J. and Brenig, (Taylor and
Francis, London, 1974) p.251

\bibitem {14}A. J. Lewis, Phys. Rev. B \textbf{13}, 2565 (1976)

\bibitem {15}G. Sherwood, M.A. Howson and G. J. Morgan; J. Phys.: Condens.
Matter \textbf{3,} 9395 (1991)

\bibitem {16}T. Kawahara, S. Tamura, H. Inai, Y. Okamoto, and J. Morimoto: J.
Mater. Res., Vol. \textbf{14,} No. 4, (1999)

\bibitem {17}S. Nakamae, D. Colson, A. Forget, I. Legros, J.F. Marucco, C.
Ayache and M. Ocio, Phys. Rev. B \textbf{63,} 092407 (2001)

\bibitem {18}A. A. Taskin, A. N. Lavrov, and Yoichi Ando; Phys. Rev. B
\textbf{73,} 121101(R) (2006)

\bibitem {19}G. S. Nolas, M. Beekman, and R. W. Ertenberg: J. Appl. Phys.
\textbf{100,} 036101 (2006)

\bibitem {20}J. J. Neumeier and H. Terashita Phys. Rev. B \textbf{70,} 214435 (2004)

\bibitem {21}J.M. Ziman, \textit{Principles of the theory of solids},
(University Press, Cambridge 1972)

\bibitem {22}Cutler, M. and Mott, N. F., Phys. Rev. \textbf{181,} 1336 (1969).

\bibitem {23}G. P. Triberis and L. R. Friedman, Journ. Non-cryst. Sol.
\textbf{79,} 29 (1986).

%\bibitem{label}
%Text of bibliographic item


%notes:
%\bibitem{label} \note


%subbibitems:
%\begin{subbibitems}{label}
%\bibitem{label1}
%\bibitem{label2}
%If there is a note, it should come last:
%\bibitem{label3} \note
%\end{subbibitems}

\end{thebibliography}
\end{document}